\documentclass{icfi}

\begin{document}
\markboth{Rodolfo Gambini, Jorge Pullin}
{Consistent discretizations and quantum gravity}
%
\catchline{}{}{}{}{}
%
\title{Canonical quantum gravity and consistent discretizations}

\author{Rodolfo Gambini}
\address{Instituto de F\'{\i}sica, Facultad de Ciencias, Igua esq. Mataojo,
Montevideo, Uruguay}
\author{Jorge Pullin}
\address{Department of Physics and Astronomy, Louisiana State University,
Baton Rouge, LA 70803 USA}

\maketitle

%
\begin{abstract}
We review a recent proposal for the construction of a quantum theory
of the gravitational field. The proposal is based on approximating the
continuum theory by a discrete theory that has several attractive properties,
among them, the fact that in its canonical formulation it is free of
constraints. This allows to bypass many of the hard conceptual problems
of traditional canonical quantum gravity. In particular the resulting 
theory implies a fundamental mechanism for decoherence and bypasses the
black hole information paradox.
\end{abstract}

\section{Introduction}

Discretizations are very commonly used as a tool to treat field
theories.  Classically, when one wishes to solve the equations of a
theory on a computer, one replaces the continuum equations by discrete
approximations to be solved numerically. At the level of quantization,
lattices have been used to regularize the infinities that plague field
theories. This has been a very successful approach for treating
Yang--Mills theories. The current approaches to non-perturbatively
construct in detail a mathematically well defined theory of quantum
gravity both at the canonical level \cite{Th} and at the path integral level
\cite{Pe} resort to discretizations to regularize the theory.

Discretizing general relativity is more subtle than what one initially
thinks. Consider a $3+1$ decomposition of the Einstein equations. One
has twelve variables to solve for (the six components of the spatial
metric and the six components of the extrinsic curvature). Yet, there
are {\em sixteen} equations to be solved, six evolution equations for
the metric, six for the extrinsic curvature and four constraints. In
the continuum, we know that these sixteen equations are {\em
  compatible}, i.e. one can find twelve functions that satisfy them.
However, when one discretizes the equations, the resulting system of
algebraic equations is in general incompatible. This is well known,
for instance, in numerical relativity \cite{Chop}. The usual attitude
there is to ignore the constraints and solve the twelve evolution
equations (this scheme is called ``free evolution''). The expectation
is that in the limit in which the lattice is infinitely refined, the
constraints will also be satisfied if one satisfied them initially.
The situation is more involved if one is interested in discretizing
the theory in order to quantize it.  There, one needs to take into
account all equations. In particular, in the continuum the constraints
form an algebra. If one discretizes the theory the discrete version of
the constraints will in many instances fail to close an algebra.
Theories with constraints that do not form algebras imply the
existence of more constraints which usually makes them inconsistent.
For instance, it might be the case that there are no wavefunctions
that can be annihilated simultaneously by all constraints. One can ask
the question if this is not happening in the construction that
Thiemann works out. To our knowledge, this issue has not been probed.
What is clear, is that discretizing relativity in order to quantize it
will require some further thinking.

The new proposal we have put forward \cite{GaPuprl}, called {\em
consistent discretization} is that, in order to make the discrete
equations consistent, the lapse and the shift need to be considered as
some of the variables to be solved for.  Then one has 16 equations and
16 unknowns. This might appear surprising since our intuition from the
continuum is that the lapse and the shift are freely specifiable. But
we need to acknowledge that the discrete theory {\em is a different
theory}, which may approximate the continuum theory in some
circumstances, but nevertheless is different and may have important
differences even at the conceptual level. This is true of any discretization
proposal, not only ours.

We have constructed a canonical approach for theories discretized in
the consistent scheme \cite{DiGaPu}. The basic idea is that one does
not construct a Legendre transform and a Hamiltonian starting from the
discretized Lagrangian picture. The reason for this is that the
Hamiltonian is a generator of infinitesimal time evolutions, and in a
discrete theory, there is no concept of infinitesimal. What plays the
role of a Hamiltonian is a canonical transformation that implements
the finite time evolution from discrete instant $n$ to $n+1$. The
canonical transformation is generated by the Lagrangian viewed as a
type I canonical transformation generating functional. The theory is
then quantized by implementing the canonical transformation as a
unitary evolution operator. A discussion of an extension of the Dirac
procedure to these kinds of systems can be seen in \cite{DiGaPoPu}.

\section{Examples}

We have applied this discretization scheme to perform a discretization
(at a classical level) of BF theory and Yang--Mills theories
\cite{DiGaPu}. In the case of BF theories this provides the first
direct discretization scheme on a lattice that is known for such
theories. In the case of Yang--Mills theories it reproduces known
results. We have also studied the application of the discretization
scheme in simple cosmological models. We find that the discretized
models approximate general relativity well and avoid the singularity
\cite{cosmo}. More interestingly, they may provide a mechanism for
explaining the value of fundamental constants \cite{gapusmolin}. When
the discrete models tunnel through the singularity, the value of the
lapse gets modified and therefore the ``lattice spacing'' before and
after is different. Since in lattice gauge theories the spacing is
related to the ``dressed'' values of the fundamental constants, this
provides a mechanism for fundamental constants to change when
tunneling through a singularity, as required in Smolin's \cite{smolin}
``life of the cosmos'' scenario.

It is quite remarkable that the discrete models work at all. When one
solves for the lapse and the shift one is solving non-linear coupled
algebraic equations. It could have happened that the solutions were
complex. It could have happened that there were many possible
``branches'' of solutions. It could have happened that the lapse
turned negative. Although all these situations are possible given
certain choices of initial data, it is remarkable that it appears that
one can choose initial data and a convenient ``branch'' of solutions
for which pathologies are avoided and the discrete theory approximates
the continuum theory in a controlled fashion. For simple cosmological
examples, the quantization implementing the evolution as a unitary
operation has been worked out in detail \cite{cosmo}.

We are currently exploring the Gowdy models with this approach,
initially at a classical level only. Here the problem is considerably
more complex than in cosmological models. The equations to be solved
for the lapse and the shift become a coupled system that couples all
points in the spatial discretization of the lattice. The problem can
only be treated numerically. Moreover, Gowdy models have a global
constraint due to the topology that needs special treatment. We have
written a fortran code to solve the system using iterative techniques
(considerable care needs to be exercised since the system becomes
almost singular at certain points in phase space) and results are
encouraging. In the end the credibility of the whole approach will
hinge upon us producing several examples of situations of interest
where the discrete theories approximate continuum GR well.

\section{Several conceptual advantages}

The fact that in the consistent discrete theories one solves the
constraints to determine the value of the Lagrange multipliers has
rather remarkable implications. The presence of the constraints is one
of the most significant sources of conceptual problems in canonical
quantum gravity. The fact that we approximate the continuum theory
(which has constraints) with a discrete theory that is constraint free
allows us to bypass in the discrete theory many of the conceptual
problems of canonical quantum gravity.

The reader may ask how is such an approximation possible. After all,
if the continuum theory has constraints and the discrete version does
not, the two theories do not even have the same number of degrees of
freedom.  This is true. What is happening is that in the discrete
theory there will generically be several solutions that approximate a
given solution of the continuum theory. As solutions of the discrete
theory they are all different yet they represent the same solution in
the continuum. Therefore it is not surprising that the discrete theory
has more degrees of freedom.

One of the main problems we can deal with due to the lack of
constraints is the ``problem of time''. This problem has generated a
large amount of controversy and has several aspects to it. We cannot
cover everything here, the definitive treatise on the subject is the
paper by Kucha\v{r} \cite{Kuchar}.

To simplify the discussion of the problem of time, let us consider an
aspect of quantum mechanics that most people find unsatisfactory
perhaps from the first time they encounter the theory as
undergraduates. It is the fact that in the Schr\"odinger equation, the
variables ``$x$'' and ``$t$'' play very different roles. The variable
$x$ is a quantum operator of which we can, for instance, compute its
expectation value, or its uncertainty. In contrast $t$ is assumed to
be a continuous external parameter. One is expected to have a clock
that behaves perfectly classically and is completely external to the
system under study. Of course, such a construction can only be an
approximation. There is no such thing as a perfect classical clock and
in many circumstances (for instance quantum cosmology) there is no
``external clock'' to the system of interest. How is one to do quantum
mechanics in such circumstances? The answer is: ``relationally''. One
could envision promoting {\em all} variables of a system to quantum
operators, and choosing one of them to play the role of a
``clock''. Say we call such variable $t$ (it could be, for instance
the angular position of the hands of a real clock, or it could be
something else). One could then compute conditional probabilities for
other variables to take certain values $x_0$ when the ``clock''
variable takes the value $t_0$. If the variable we chose as our
``clock'' does correspond to a variable that is behaving classically
as a clock, then the conditional probabilities will approximate well
the probabilities computed in the ordinary Schr\"odinger theory. If
one picked a ``crazy time'' then the conditional probabilities are
still well defined, but they don't approximate any Schr\"odinger
theory well. If there is no variable that can be considered a good
classical clock, Schr\"odinger's quantum mechanics does
not make sense and the relational quantum mechanics is therefore
a generalization of Schr\"odinger's quantum mechanics.

The introduction of a relational time in quantum mechanics therefore
appears well suited as a technique to use in quantizing general
relativity, particularly in cosmological situations where there is no
externally defined ``classical time''. Page and Wootters advocated
this in the 1980's \cite{PaWo}.  Unfortunately, there are technical
problems when one attempts the construction in detail for general
relativity. The problem arises when one wishes to promote the
variables to quantum operators. Which variables to choose? In
principle, the only variables that make sense physically are those
that have vanishing Poisson brackets (or quantum mechanically
vanishing commutators) with the constraints. But since the Hamiltonian
is one of the constraints, then such variables are ``perennials'' i.e.
constants of motion, and one cannot reasonably expect any of them to
play the role of a ``clock''. One could avoid this problem by
considering variables that do not have vanishing Poisson brackets with
the constraints. But this causes problems. Quantum mechanically one
wishes to consider quantum states that are annihilated by the
constraints. Variables that do not commute with the constraints as
quantum operators map out of the space of states that solve the
constraints. The end result of this, as discussed in detail by
Kucha\v{r} \cite{Kuchar} is that the propagators constructed with the
relational approach do not propagate.

Notice that all the problems are due to the presence of the
constraints. In our discrete theory, since there are no constraints,
there is no obstruction to constructing the relational picture. We
have discussed this in detail in \cite{greece}. 

Of great interest is the fact that the resulting relational theory
will never entirely coincide with a Schr\"odinger picture. In
particular, since no clock is perfectly classical, pure states do not
remain pure forever in this quantization, but slowly decohere into
mixed states. We have estimated the magnitude of this effect.
In order to do this, we chose the ``best possible classical clock'' as 
constructed by Ng and Van Damme and Amelino-Camelia \cite{AcNg}
elaborating on the
pioneering work of Salecker and Wigner \cite{Wigner}. The result is that the 
rate of decoherence 
is proportional to $\omega^2 T^{4/3}_{\rm Planck} T^{2/3}$ where $
\omega$ is the frequency associated with the spread in energy 
levels of the system under study, $T_{\rm Planck}$ is Planck's time
and $T$ is the time that the system lives. The effect is very
small. Only for systems that have rather large energy spreads (
Bose--Einstein condensates are a possible example) the effect may be close
to observability. With current technologies, the condensates do not
have enough atoms to achieve the energy spreads of interest, but it
might not be unfeasible as technology improves to observe the effect
\cite{deco}.

The fact that a pure state evolves into a mixed state opens other
interesting possibilities, connected with the black hole information
puzzle. This puzzle is related to the fact that one could consider a
pure quantum state that collapses into a black hole. The latter will
start evaporating due to Hawking radiation until eventually it
disappears. What one is left with at the end of the day appears to be
the outgoing radiation, which is in a mixed state. Therefore a pure
state appears to have evolved into a mixed state. There is a vast
literature discussing this issue (see for instance \cite{GiTh} for a
short review). Possible solutions proposed include that the black hole
may not disappear entirely or that some mechanism may allow pure
states to evolve into a mixed state. But we have just discussed that
the relational discrete quantum gravity predicts such decoherence! We
have estimated that the decoherence is fast enough to turn the pure
state into a mixed one before the black hole can evaporate completely,
\cite{infopuzzle2} (an earlier manuscript we wrote did not use the
optimal clocks and the rate of decoherence it predicted was not as
decisive \cite{infopuzzle1}). The result is quite remarkable, since
the decoherence effect, as we pointed before, is quite small. It is
large enough to avoid the information puzzle in black holes, even if
one considers smaller and smaller black holes which evaporate faster
since they also have larger energy spreads and therefore the
decoherence effect operates faster.

\section{Summary}

Analyzing in detail how to discretize general relativity led us to
develop a way to consistently discretize the theory, in the sense that
all the resulting discrete equations can be solved simultaneously.
Surprisingly, the consistent discretizations not only approximate
general relativity well in several situations, but as theories are
conceptually much simpler to analyze than the continuum theory, since
they do not have constraints.  This allows us to handle several of the
hard conceptual problems of canonical quantum gravity.  What is now
needed is to demonstrate that the range of situations in which the
discrete theory approximates general relativity well is convincingly
large enough to consider its quantization as a route for the
quantization of general relativity.

\section{Acknowledgments}

This work was
supported by grant NSF-PHY0244335 and funds from the Horace Hearne
Jr. Institute for Theoretical Physics.

\end{document}